\documentclass{article}
\usepackage[utf8]{inputenc}
\usepackage{authblk}

\title {Demons in Black Hole Thermodynamics: Bekenstein and Hawking}
\author{Galina Weinstein\thanks{This work is supported by ERC advanced grant number 834735.}}
\affil{\normalsize The Department of Philosophy, University of Haifa, Haifa, the Interdisciplinary Center (IDC), Herzliya, Israel.} 

\date{March 3, 2021}

\begin{document}

\maketitle

\begin{abstract} 
This paper comprehensively explores Stephen Hawking’s interaction with Jacob Bekenstein. Both Hawking and Bekenstein benefited from the interaction with each other. It is shown that Hawking's interaction with Bekenstein drove him to solve the problems in Bekenstein's black hole thermodynamics in terms of a new thermal radiation mechanism. Hawking argued that Bekenstein’s thermodynamics of black holes was riddled with problems. In trying to solve these problems, Hawking cut the Gordian knot with a single stroke: black holes emit thermal radiation. Hawking derived the thermal radiation using a semi-classical approximation in which the matter fields are treated quantum mechanically on a classical space-time background. Hawking's semi-classical approximation yielded a simple formula for the entropy of the black hole, which turned out to be equivalent to Bekenstein’s equation for entropy. 

\end{abstract}

\section{Preface}

This paper comprehensively explores Stephen Hawking’s interaction with Jacob Bekenstein. I show in this paper that both Hawking and Bekenstein benefited from the interaction with each other. 

Bekenstein ascribed entropy to black holes and suggested that the entropy of black holes is proportional to the surface area of the event horizon of the black hole (Section \ref{Bekent}). Hawking proposed an area theorem, which suggested to Bekenstein that the black hole entropy should tend to grow (Section \ref{area}). 

It is shown in this paper that Hawking's interaction with Bekenstein drove the former to solve the problems in Bekenstein's black hole thermodynamics in terms of a new thermal radiation mechanism (Section \ref{radiation}). 

Hawking argued that Bekenstein’s thermodynamics of black holes was riddled with problems. There were two main paradoxes in black hole thermodynamics that awaited solutions: Geroch's engine (Section \ref{Geroch}) and the black hole in colder radiation bath problem (Section \ref{bath}). I thoroughly analyze these paradoxes (Sections \ref{bound} and \ref{GSL2}). In trying to solve these problems, Hawking cut the Gordian knot with a single stroke: black holes emit thermal radiation. 

Hawking was afraid that if Bekenstein found out about his discovery, Bekenstein would use it as a further argument to support his entropy of black holes, which Hawking did not like. 

Hawking derived the thermal radiation using a semi-classical approximation in which the matter fields are treated quantum mechanically on a classical space-time background. 
What persuaded him that the thermal radiation from black holes was real, was that it was exactly what was required to identify the surface area of the event horizon, with the entropy of the black hole (Sections \ref{semi} and \ref{entropy}). Hawking's semi-classical approximation yielded a "simple formula" for the entropy of the black hole. It turned out to be equivalent to Bekenstein’s equation for the entropy. But Hawking fixed Bekenstein’s constant of proportionality in the equation for entropy. 

On his 60th birthday, Hawking said: "I would like this simple formula to be on my tombstone" \cite{Hawking 2003}, p. 113 (Section \ref{entropy}). 

\section{Black holes have no hair} \label{hair}

In spring 1922, Albert Einstein gave lectures and discussion sessions at the Collège de France in Paris. In the second discussion session, Jacques Hadamard, a celestial mechanics professor at the Collège de France, posed before Einstein a query: Since there is a singularity at $r$ = 0 in the Schwarzschild solution [see equation (\ref{eq22}) in Section \ref{Geroch}], one may ask, what is the physical meaning of the Schwarzschild solution? We may ask whether $r$ = 0 practically and physically in our world? This question embarrassed Einstein. He said that if the radius term could really become zero somewhere in the universe, then it would be an unimaginable disaster for his general theory of relativity. Einstein considered that it would be a catastrophe, and jokingly called it the "Hadamard catastrophe” \cite{Nordmann}, pp. 155-156.

Fast forward, Roger Penrose wrote: "We may ask whether any connection is to be expected between the existence of a trapped surface and the presence of a physical space-time singularity such as that occurring at $r$ = 0 in” the Schwarzschild solution? Penrose supplied the answer by a singularity theorem he had proved together with Stephen Hawking: "the presence of a trapped surface always does imply the presence of some form of space-time singularity”. That is, a singularity must arise when a stellar-mass black hole is formed \cite{Penrose}
pp. 1151-1152; \cite{Hawking and Penrose}, pp. 538-539.

Brandon Carter and Werner Israel obtained uniqueness theorems. Israel provided a theorem which states that the Schwarzschild and Reissner-Nordström solutions are the only static solutions with an absolute event horizon \cite{Israel}, p. 164. Carter established the family of Kerr solutions as the general pseudo-stationary, asymptotically flat, vacuum solutions with an absolute event horizon \cite{Carter}, p. 332. This is the Israel Carter Conjecture. The following picture then suggested itself to Penrose. A certain body collapses to the size of the gravitational radius after which a trapped surface is found in the region surrounding the matter. Outside the trapped surface there is a surface which is the absolute event horizon. "Ultimately the field settles down to becoming a Kerr solution (in the vacuum case)” or a Kerr-Newman solution if a nonzero net charge is trapped in the black hole \cite{Penrose}, pp. 1156-1157). 

The above theorem which Hawking called the Carter-Israel conjecture \cite{Hawking 1971}, p. 1344, states that "if an absolute event horizon develops in an asymptotically flat space-time, then the solution exterior to this horizon approaches a Kerr-Newman solution asymptotically with time”. Moreover, the Kerr and Kerr-Newman solutions are explicit asymptotically flat stationary solutions of the field equations of general relativity involving just three free parameters: mass, angular momentum and charge. That is, "Only the mass, angular momentum and charge need survive as ultimate independent parameters”\cite{Penrose}, p. 1158. 

The Carter-Israel conjecture was embellished by John Archibald Wheeler who famously said: "a black hole has no hair”. Hair refers to all the parameters which are hidden to outside observers. According to Wheeler, all the physical properties of the infalling matter into a black hole are eliminated and we cannot distinguish between two black holes with the same mass, angular momentum (spin) and electrical charge \cite{Wheeler 1981}, pp. 32-33. 

\section{Bekenstein exorcises Wheeler’s demon} \label{demon}

During 1971, Wheeler was pondering about things falling into a black hole and having everything disappear there. He thought to himself that there are some ramifications that are not immediately obvious. He brooded about the second law of thermodynamics that says that everything degenerates to a uniform temperature and imagined the following irreversible process: suppose you put a hot tea cup next to a cold tea cup. They come into a common temperature. By putting these two together, we have contributed to the degree of the entropy of the universe and its information loss. 

One day when he was still a graduate student at Princeton, Jacob Bekenstein entered Wheeler’s office and Wheeler told him: "Jacob, if a black hole comes by, I can drop both tea cups into the black hole and conceal the evidence of my crime”. Wheeler’s demon, as Bekenstein called it, committed the perfect crime against the second law of thermodynamics. That is because when the tea cups are dropped into the black hole, entropy is destroyed inside the black hole. According to the no-hair theorem, no one outside the black hole would ever notice any change in the black hole because all the physical properties of the infalling matter into the black hole are eliminated.

After mulling over this problem, Bekenstein came back to Wheeler in a few months. In the meeting between the two, Bekenstein told Wheeler that the demon had not avoided the entropy increase: "you’ve just put the entropy increase in another place. The black hole itself already has entropy and you simply increase it” \cite{Bekenstein 1980}, pp. 24-25; \cite{Wheeler 2008}.

\section{Hawking’s area theorem and the the Penrose Process}  \label{area}

In 1971 Hawking formulated an area theorem: the area of the black hole never decreases. It increases for all black holes. Hawking considered a situation in which there are initially two collapsed objects or black holes a considerable distance apart. The black holes are assumed to have formed at some earlier time as a result of either gravitational collapses or, the amalgamation of smaller black holes. Suppose now that the two black holes merge to form a single black hole which settles down to a Kerr black hole with mass $M_{{BH}}$ and spin $a$. The area of the event horizon of the resulting black hole is greater than the sum of the areas of the event horizons around the original black holes: $A_K\geq(A_1+A_2)$.

In 1969 Penrose had written: "I want to consider the question of whether it is possible to extract energy out of a 'black hole'”. Penrose suggested mechanisms for extracting energy from a black hole \cite{Penrose}, p. 1159. 

Two years later Penrose and R. Floyd published a paper on the extraction of rotational energy from black holes which became to be known as the Penrose process. Penrose and Floyd discuss the question whether the mass-energy of a rotating Kerr black hole could, under certain circumstances, be a source of available energy. In order to answer this question they consider a process of an extraction of rotational energy from a Kerr black hole. 

Penrose and Floyd explain that there are two regions lying outside of the horizon of a Kerr black hole: first, the stationary limit is a surface at which a particle would have to travel with a velocity of light in order to appear stationary to an observer at infinity. The stationary limit lies outside the event horizon and in between there is region called the ergosphere. A particle can enter and leave the ergosphere but as viewed from infinity inside, it cannot remain stationary. The absolute event horizon is the effective boundary of the black hole inside which no information can escape to the outside world. The stationary limit and the event horizon coincide in the case of zero angular momentum, i.e. for a stationary Schwarzschild black hole.  

Penrose and Floyd consider the Killing vector $K= \frac {\partial}{\partial t}$. $K$ becomes null at the stationary limit and space-like within the ergosphere. Penrose and Floyd use $K$ to define the energy of a particle with four-momentum (as measured from infinity) $p_0$: $E=-Kp_0$. Since $K$ is space-like within the ergosphere, it is possible for $E$ to be negative there, even though the four-momentum is time-like.    

Suppose a particle with four-momentum $p_a$ enters the ergosphere. It then splits into two. According to conservation of momentum: $p_{a0} =p_{a1} + p_{a2}$, and according to conservation of energy: $E_{a0} =E_{a1} + E_{a2}$. So that, the mass-energy (as measured from infinity) of the first particle is negative $E_{a1} < 0$ and it will cross the event horizon and fall into the black hole whilst the second, with positive energy (as measured from infinity) $E_{a2} > 0$, escapes back to infinity and carries more mass-energy than the original particle possessed: $E_{a2} > E_{a0}$ \cite{Penrose2}, pp. 177-178.  

According to the Carter-Israel no-hair theorem, the area of the event horizon of the newly formed Kerr-Newman black hole is given by: 

\begin{equation} \label{eq 25}
A=8\pi M_{BH} r_+,
\end{equation}

\vspace{1mm} 
\noindent where $r_+$ is the location of the surface of the horizon:
\vspace{1mm} 

\begin{equation} \label{eq 26}
r_+=M_{BH}+\sqrt{{M_{BH}}^2-\frac{J}{M_{BH}}-Q^2},
\end{equation}
\vspace{1mm} 

\noindent and $J$ stands for the angular momentum. For a Kerr black hole $Q$ = 0 \cite{Hawking 1971}, p. 1345; \cite{Hawking 1977}, p. 35.

For the particle that falls into the black hole Penrose and Floyd wrote an inequality equation: $2M_{BH}r_+E_{a1} -\Omega \Lambda_1\geq 0$, 
\vspace{1mm} 
where $\Omega$ is the angular velocity of the black hole, and $\Lambda_1$ is the angular momentum of the first particle. The mass of the black hole decreases: $dM_{BH}=-E_{a1}$ and the angular momentum of the black hole also changes as a result of the rotational energy that is extracted: $dJ_1=-\Lambda_1$. Consequently, $dA \geq 0$ [see equation (\ref{eq40})]. 

The size of the black hole (as measured by $A$) increases even though its mass can decrease. We thus retrieve the area theorem. Penrose and Floyd conclude: “In fact, from general considerations one may infer that there should be a natural tendency for the surface area of the event horizon of a black hole to increase with time whenever the situation is non-stationary. Thus the ideal of maximum efficiency would appear to be achieved whenever this surface area increase is as small as possible” (i.e. a reversible process) \cite{Penrose2}, pp. 177-178.

At the end of the paper the authors "thank R. P. Geroch for discussions". The paper was received in Dcember 1970. In December 1971 Robert Geroch would present his thought experiment: the Geroch heat engine (see Section \ref{Geroch}) \cite{Penrose2}, p. 179. 

\section{Wheeler's students}  \label{students}

In 1970 Demetrios Christodoulou, then a graduate student of Wheeler’s, introduced the concept of "irreducible mass", $m_{ir}$, of the black hole and considered two kinds of transformations: reversible and irreversible. He arrived at the following formula:

\begin{equation} \label{eq1}
m^2=m_{ir}^2+ \frac{J^2}{4m_{ir}^2},
\end{equation}

\noindent where $G=c=1$ and $J$ is the angular momentum of the black hole. This equation is equivalent to $E^2=m^2+p^2.$

Equation (\ref{eq1}) says that the total mass-energy of the black hole is equal to $m_{ir}$ plus a rotational energy term, which can be changed by the extraction of energy from the black hole.

In addition to writing equation (\ref{eq1}), Christodoulou imposed four conditions: 
\vspace{1mm} 

1) One can approach arbitrarily closely to reversible transformations that augment or deplete the rotational contribution: 
\vspace{1mm} 
\noindent $\frac{J^2}{4m_{ir}^2}$ to $m^2$.

2) The attainable range of reversible transformation extends from: 
\vspace{1mm} 

$L=0, m^2=m_{ir}^2$ to $L=m^2, m^2=2m_{ir}^2.$
\vspace{1mm} 

3) An irreversible transformation is characterized by an increase in $m_{ir}$.  

4) "There exists no process which will decrease the irreducible mass". 
\vspace{1mm} 

Christodoulou examined the Penrose process and demonstrated that if condition 2) is fulfilled, then equation (\ref{eq1}) applies to the particle that falls into the black hole. 

In other words, the most efficient processes are those associated with reversible changes of the black hole (reversible transformations). The irreversible transformation which is characterized by an increase in $m_{ir}$ of the black hole was found to be less efficient \cite{Bekenstein 1980}, p. 25. 

Bekenstein and Christodoulou recognized that for a Schwarzschild black hole:

$m_{ir}^2 \propto A/16 \Pi.$ 
\vspace{1mm} 

\noindent [see equation (\ref{eq6})] \cite{Christodoulou}, pp. 1596-1597; \cite{Bekenstein 1973}, p. 2333; \cite{Bekenstein 1980}, p. 25. 

Moreover, in thermodynamics, reversible processes are the most efficient ones for converting energy from one form to another. So, Bekenstein thought to himself that black holes must be in harmony with thermodynamics \cite{Bekenstein 1980}, p. 25. 

\section{Bekenstein’s entropy} \label{Bekent}

Bekenstein had a sneaking suspicion that the area $A$ of the black hole "might play a role of the entropy of black holes” \cite{Bekenstein 1980}, p. 25. 

In 1972 Bekenstein ascribed entropy to black holes. In fact, Bekenstein had already suggested in his Ph.D thesis that the entropy of black holes is proportional to the surface area of the event horizon of the black hole $A$:  

\begin{equation} \label{eq2}
S_{BH}=\frac{\eta kA}{l^2_p},
\end{equation}

\noindent where $\eta$ is a constant. On Wheeler’s suggestion Bekenstein took the proportionality constant of the order $\frac{1}{l_p^2}$, 
\vspace{1mm} 
where $l_p$ is the Planck length [see equation (\ref{eq19})] \cite{Bekenstein 1972}, p. 733; \cite{Bekenstein 2001}, p. 515; see explanation in Section \ref{mini}.

Bekenstein explained why he chose equation (\ref{eq2}). Christodoulou's research in classical black hole physics and Hawking's area theorem suggested to him that the central desirable property was that black hole entropy should tend to grow. 
Equation (\ref{eq2}) complied with the area theorem because when two black holes merge to form a single black hole which settles down to a Kerr black hole, the entropy of the resulting black hole will increase. The area of the black hole (a geometric quantity) behaves very much like entropy (a thermodynamic property) \cite{Bekenstein 1972}, p. 733; \cite{Bekenstein 2001}, p. 515. 

\section{Interpretation of the entropy} \label{entropy2}

According to Bekenstein, $S_{BH}$ can be interpreted as the information hidden inside the black hole and the ignorance of an observer at infinity about the matter inside it. 

In 1948 Claude Shannon famously presented a definition of the entropy which "measures the amount of information generated by the source per symbol or per second" \cite{Shannon}, p. 396. I will not discuss Shannon's ideas further. I will just add that Shannon wrote: "From our previous discussion of entropy
as a measure of uncertainty it seems reasonable to use the conditional entropy of the message, knowing the received signal, as a measure of this missing information" \cite{Shannon}, p. 407. 
In 1973 Bekenstein wrote that "the entropy associated with the system is given by Shannon's formula". And he considered the Shannon information entropy of a system which measures our ignorance or lack of information about the actual internal configuration of the system. 

Léon Brillouin identified information with negative entropy \cite{Brillouin}, pp. IX, pp. 7-10. Bekenstein adopted Brillouin's relationship between information and negentropy and stressed that the more information entropy, the less information we have about the black hole \cite{Bekenstein 1973}, p. 2335.

Bekenstein further explained that "the entropy of an evolving thermodynamic system increases due to the gradual loss of information which is a consequence of the washing out of the effects of the initial conditions". That is, the entropy of a thermodynamic system, i.e. a black hole, which is not in equilibrium increases. That is because information about the internal configuration of the system is being lost during its evolution. This happens as a result of the wiping out of the initial conditions of the system. Bekenstein summoned Maxwell's demon, saying: "It is possible for an exterior agent to cause a decrease in the entropy of a system by first acquiring information about the internal configuration of the system. The classic example of this is that of Maxwell's demon". But in acquiring information about the system, the demon inevitably causes an increase in the entropy of the rest of the universe. Thus, even though the entropy of the system decreases, the over-all entropy of the universe increases in the process \cite{Bekenstein 1973}, p. 2336. In 1993, Bekenstein would raise a somewhat similar argument, see Section \ref{demon2}.

As a black hole approaches equilibrium, it loses its hair. Only $M$, $J$, and $Q$ are left as parameters of the black hole at later times. It is conjectured that the loss of information about the initial conditions of the black hole would be reflected in a gradual increase in $S_{BH}$. According to Hawking's area theorem, $S_{BH}$ increases monotonically as the black hole evolves \cite{Bekenstein 1973}, p. 2336. 

\section{Bekenstein’s GSL}  \label{GSL}

Entropy ebbs and flows at will and Wheeler’s demon violates the second law of thermodynamics. Entropy increases inside the black hole whilst there is a decrease of entropy outside the black hole (this is explained in detail in the next section). Hawking’s area theorem $dS_{BH}\geq0$ is contradicted. 

Bekenstein solved the paradox posed by Wheeler’s demon by suggesting that the sum of the change of entropies $dS_g$, of the black hole $dS_{BH}$ and the matter outside of the black hole $dS_M$, must never decrease ($dS_g\geq0$): 

\begin{equation} \label{eq37}
dSg=dS_{BH}+dS_M \geq 0.    
\end{equation}

\noindent He termed this, the \emph{Generalized Second Law of Thermodynamics} (hereafter GSL). 

The area theorem requires only that $A$ not decrease. But the GSL also demands that if $S_M$ is lost into black holes, $A$ will increase sufficiently for the associated increase in $S_{BH}$ to at least compensate for the decrease in $S_M$ 
\cite{Bekenstein 1972}, p. 738; \cite{Bekenstein 1973}, p. 2334-2338; \cite{Bekenstein 1975}, pp. 3077; \cite{Bekenstein 1981}, p. 287.   

\section{Geroch’s heat engine} \label{Geroch}

In December 1971, in a Colloquium at Princeton University, Geroch conceived a heat engine that employs a Schwarzschild black hole as an energy sink. Geroch's thought experiment could violate Bekenstein’s GSL. 

Consider the Schwarzschild metric:
\vspace{1mm} 

\begin{equation} \label{eq22}
ds^{2}=\left({1}-\frac{2R_g}{r}\right)c^2dt^{2}-\frac{dr^{2}}{1-\frac{2R_g}{r}}-r^{2}\left( d \theta^2 \sin^2 \theta d \phi^2 \right),
\end{equation}

\noindent where:

\begin{equation} \label{eq23}
\noindent R_g = \frac{GM_{BH}}{c^2}
\end{equation}

\noindent is the gravitational radius.
\vspace{1mm} 

Here are the steps of the process:
\vspace{1mm} 

1) We fill a box with radiation of energy $E=\left(Mc^2\right)$, temperature $T$ and high entropy $S$. We then slowly lower the box by a rope toward the horizon of the Schwarzschild black hole in its gravitational field. Let us suppose, for the sake of the argument that the box has no weight and the rope also has no mass. We want to extract work from the Geroch engine. 

2) Suppose the box is lowered as far as possible so that $r \rightarrow 2R_g$. $r$ is the radius away from the black hole to which the box has been lowered. 

The box never actually reaches $2R_g$. The small box is then opened and the radiation is allowed to escape into the black hole. 
The gravitational field does work on the box in this process:
\vspace{1mm} 

$W = \frac{2R_g}{r}{Mc^2}$,
\vspace{1mm} 

\noindent where the gravitational potential is: $U = -\frac{2R_g}{r}{Mc^2}$ and the redshift factor is:

\begin{equation} \label{29}
\alpha \approx \frac{2R_g}{r}.
\end{equation}

3) The box is then pulled up back at expense of almost no work and can be refilled with thermal radiation from the reservoir.
\vspace{1mm} 

Since $r \rightarrow 2R_g$, the amount of work that can be extracted from the Geroch heat engine is: $W = Mc^2$ and the black hole would end up in the same state in which it began and all of the energy of the radiation could be converted to work in the laboratory from which one was doing the lowering \cite{Bekenstein 1972}, p. 148, pp. 739-740. 

The Carnot efficiency of the Geroch heat engine is: $\eta_{TH} \leq {\left( 1-\frac{T_C}{T_O} \right)}$ where $T_O$ is the temperature at which the radiation entered the black hole and $T_C$ is the temperature of the black hole into which the box exhausts its radiation. If all the mass-energy is converted to work, then the efficiency $\eta_{TH}\leq{1}$ and $T_C=0$. Consequently, Geroch stressed that black holes are systems at absolute zero temperature. And this would violate Bekenstein’s GSL because entropy in the outer world would decrease $dS_M<0$, with no counter increase in entropy in the black hole $dS_{BH}=0$. Thus, $dS_g=dS_M<0$.

Bekenstein saw a resemblance between Christodoulou's work on the Penrose process (see Section \ref{students}) and “Geroch’s gedanken experiment”. Bekenstein wrote that “Christodoulou has described a process by means of which a particle can be added to a black hole without there resulting an increase in the black hole’s area. One could imagine that the particle in question carries entropy. It would then appear that this entropy goes down the black hole without there being a compensating increase in $S_{b.h.}$ as demanded by the second law. […] It follows that the primary particle is required to fall from far away to the horizon”. Bekenstein explained that in such a fall the particle will radiate into the black hole gravitational waves with energy of the order of its rest mass. This radiation will cause a non-negligible increase in the black hole area corresponding to the increase in $S_{BH}$ demanded by the GSL. In this manner Bekenstein reconciled Christodoulou’s process with the second law \cite{Bekenstein 1972}, p. 150.

He then tried to resolve the difficulty raised by the Geroch thought experiment. In an attempt to salvage his GSL, Bekenstein wrote:
\vspace{1mm} 

\begin{equation} \label{eq24}
\noindent  S_{BH}=\frac{\eta kA}{l^2_p\hbar}, 
\vspace{1mm} 
\end{equation}

\noindent where $\hbar=\frac{h}{2\pi}$.
\vspace{1mm} 

This means that the entropy of a black hole would be enormous compared with the entropy of typical ordinary matter of comparable size and energy. Bekenstein thus assumed that the system is “dictated by quantum effects” and he gave a quantum explanation for why the box would never reach $2R_g$. But in 1980 he admitted that these attempts to salvage his GSL “were forced and inelegant” \cite{Bekenstein 1974}, p. 3292, pp. 3295-3296, p. 3298; \cite{Bekenstein 1980}, pp. 27-28; \cite{Wald}, pp. 5-6; \cite{Unruh2}, p. 154.  

Hawking objected to Bekenstein's entropy, equation (\ref{eq2}). According to Kip Thorne, "Bekenstein was not convinced. 
All the world's black hole experts lined up on Hawking's side – all, that is, except Bekenstein's mentor, John Wheeler. 'Your idea is just crazy enough that it might be right', Wheeler told Bekenstein". As said above, Bekenstein suggested that the black hole must have enormous amount of entropy [equation (\ref{eq24})] and he interpreted equation (\ref{eq2}) in terms of an "internal configuration" of the black hole, see Section \ref{entropy2}. "Nonsense, responded most of the leading black-hole physicists, including Hawking and me. The hole's interior contains a singularity, not atoms or molecules", so writes Thorne \cite{Thorne}, pp. 425-426.  

\section{The Hawking, Bardeen and Carter four laws of black hole mechanics}

Now there was a little dispute between Hawking and Bekenstein lurking beneath the surface. “I remember”, said Wheeler, “when Bekenstein’s paper came out the whole argument seemed so implausible to Stephen Hawking and his friend that they decided to write a paper to prove that it was wrong” \cite{Wheeler 2008}. Indeed, in 1973 Hawking, Bardeen and Carter, wrote: “The fact that the effective temperature of a black hole is zero means that one can in principle add entropy to a black hole without changing it in any way. In this sense a black hole can be said to transcend the second law of thermodynamics”. That is, the second law is made irrelevant \cite{BCH}, pp. 168-169.

Bardeen, Carter and Hawking formulated the four laws of black hole mechanics, which correspond to the ordinary three laws of thermodynamics and the zeroth law of thermodynamics. Recall that according to the no-hair theorem, when an object collapses to create a black hole, it will result in a stationary state that is characterized by only three parameters: $M_{BH}$, $J$, and $Q$. Apart from these three properties the black hole apparently preserves no other information of the object that has collapsed. Now suppose that radiation falls onto the black hole, these are the four laws of black hole mechanics \cite{BCH} , pp. 167-168: 

0) The surface gravity $\kappa$ is constant over the event horizon of a stationary black hole.

1) The first law relates the change of the mass, angular momentum and charge of the black hole to the change of its entropy:

\begin{equation} \label{eq3}
d\left(M_{BH} c^2 \right)= T_{BH} dS_{BH} -\Omega dJ- \Phi dQ,
\end{equation}

\noindent where $d\left(M_{BH} c^2 \right)$ is the corresponding change in the black hole energy, $\Omega$ stands for the angular velocity of the black hole, $J$ is the angular momentum, $\Phi$ represents the electric potential and $Q$ is the electric charge. $\Omega dJ$ and $\Phi dQ$ are analogous to $PdV$ of the first law of thermodynamics. So, $\Omega dJ$ and $\Phi dQ$ represent the work done on the black hole by adding to it angular momentum and charge, respectively. As seen from equation (\ref{eq3}), the first law is related to the no-hair theorem.

2) Hawking’s 1971 area theorem: 

\begin{equation} \label{eq40}
dA \geq 0.    
\end{equation}

\noindent Hence, $A$ can only increase and $dS_{BH}\geq 0$.

3) It is impossible by any procedure to reduce $\kappa$ to zero in a finite number of steps.

Robert Wald writes: “The fact that black holes obey such laws was, in some sense, supportive of Bekenstein’s thermodynamic ideas” \cite{Wald}, p. 6. 

\section{Bekenstein's black hole temperature}

Bekenstein responded to Bardeen, Carter and Hawking’s paper by assigning to the black hole a finite non-zero temperature that corresponds to the entropy. He wrote in 1973 equation (\ref{eq3}) as follows: 

\begin{equation} \label{eq4} d\left(M_{BH} \right)= \frac{\kappa c^3}{8 \pi G} dA - \Omega dJ- \Phi dQ,\end{equation}

\noindent where $\kappa$ is given by equation (\ref{eq7}) below. Hawking also wrote equation (\ref{eq4}) in 1973 \cite{Hawking 1973}, p. 270.

According to equations (\ref{eq3}) and (\ref{eq4}), if $G=c=1$, $T_{BH} dS_{BH} = \left(\frac{\kappa}{8 \pi}\right) dA.$ 
\vspace{1mm} 
Using the above form of the first law of black hole mechanics and equation (\ref{eq2}), Bekenstein wrote \cite{Bekenstein 1973}, p. 2335, p. 2338: 

\begin{equation} \label{eq5}
T_{BH}=\frac{\kappa l^2_p}{\eta k}.\end{equation}

Later Bekenstein explained that equation (\ref{eq2}) "incidentally, established the black hole
temperature [...] What this temperature
meant operationally was not clear, though I did study the matter in detail" \cite{Bekenstein 2001}, p. 517.

In 1975 Hawking stressed that Bekenstein "was the first to suggest that some multiple of $\kappa$ should be regarded as representing in some sense the temperature of black holes. He also pointed out that one has the relation" equation (\ref{eq4}) \cite{Hawking 1975}, p. 192. 

\section{A black hole in a thermal radiation bath} \label{bath}

Bekenstein's GSL started to wobble slightly as it faced
another thought experiment. Consider a Kerr black hole of mass $M_{BH}$ immersed in a black body radiation bath within a cavity of temperature $T$. We assume that the cavity does not rotate so that the radiation has zero angular momentum and that the process of accretion has reached a steady state. Then to radiation of energy $E$ (as measured from infinity) going down the black hole there corresponds $dS_{BH} = \frac{E}{T_{BH}}$ [see also equation (\ref{eq17}) Section \ref{GSL2}].

Bekenstein explained that since the accretion onto the black hole takes place in steady state, then together with $E$, entropy $S=\frac{E}{T}$ flows into the black hole. And so, for $T \geq T_{BH}$ we get $dS_g =dS_{BH} - S \geq 0$ and the GSL is not violated. For $T < T_{BH}$, however, the GSL is violated. 

In order to solve this problem, Bekenstein suggested that the wavelength in the radiation $\frac{\hbar}{T}$ is already larger than the size of the black hole (very long wavelength corresponding to low $T$). The radiation no longer flows in as a continuous fluid would, but rather most wavelengths in it must tunnel into the black hole discretely and individually. Bekenstein argued that the rate of tunnelling is sensitive to the wavelength. The shorter the wavelength, the higher the rate. Thus for $T \geq T_{BH}$ the rate is high and for $T < T_{BH}$, the rate is so slow that practically the radiation can hardly flow into the black hole. 

For $T < T_{BH}$, Bekenstein further introduced an ad hoc hypothesis that irreversible processes will go on in the cavity material (and perhaps in the radiation also), which will generate additional entropy. This extra entropy suffices to make the GSL work in the regime $T < T_{BH}$ \cite{Bekenstein 1974}, p. 3298. This explanation seemed cumbersome and it did not comply with Occam’s razor.

\section{Making order of the pandemonium} \label{order}

Bekenstein suggested on thermodynamic grounds that a black hole should have a finite entropy $S_{BH}$ that is proportional to the surface area of the event horizon $A$ [see equation (\ref{eq2})]. He further proposed that the temperature of the black hole should be regarded as proportional to the surface gravity $\kappa$, see equations (\ref{eq5}) and (\ref{eq7}) below \cite{Hawking 1974} p. 31. 

Hawking reasoned that if the black hole has a finite entropy that is proportional to the area of the event horizon $A$:

\begin{equation} \label{eq6}
 A = 4\pi \left(2{R_g}\right)^2 = 16 \pi \frac{\left(G M_{BH}\right)^2}{c^4}, \end{equation}

\noindent it must necessarily have a finite temperature $T_{BH}$ that is proportional to the surface gravity of the black hole $\kappa$:

\begin{equation} \label{eq7} \kappa = \frac{G M_{BH}}{\left(2{R_g}\right)^2} = \frac{c^4}{4G M_{BH}}. \end{equation}

\noindent $R_g$ is defined by equation (\ref{eq23}).

Hawking finally argued that this in turn indicates that a black hole can be in equilibrium with thermal radiation at some temperature other than zero. It is certainly true for bodies in classical physics, but it is not so for black holes which can devour matter and cannot emit anything \cite{Hawking 1977}, p. 37. 

In 1973 Hawking, Bardeen and Carter, thought that Bekenstein's resolution to the paradox of the black hole-in-thermal-radiation-bath was far-fetched (see Section \ref{bath}). A black hole, said the three authors, cannot be in equilibrium with black body radiation at a non-zero temperature. That is because no radiation could be emitted from the black hole whereas some radiation would always cross the horizon into the black hole. If the wavelength of the radiation were very long, as suggested by Bekenstein, corresponding to a low black body temperature, the rate of absorption of radiation would be very slow. True equilibrium, however, would be possible only if there were no radiation present at all; that is, if the external black body radiation temperature were zero \cite{BCH}, p. 169.

So, Hawking was baffled about Bekenstein’s efforts at salvaging his GSL. Which is why Hawking thought that no equilibrium of a black hole with thermal radiation at some temperature other than zero was possible because the black hole would absorb any thermal radiation that fell onto it but would be unable to emit anything in return \cite{Hawking 1977}, p. 37.

The problem Hawking was faced with was: either the temperature $T_{BH}$ is identically zero, in which case $S_{BH}$ [equation (\ref{eq2})] is infinite and the concept of black hole entropy is meaningless, or black holes have to emit thermal radiation with some finite nonzero temperature \cite{Hawking 1976}, p. 193. 

"The first case", says Hawking, "is what holds in purely classical theory, in which black holes can absorb but do not emit anything. Bekenstein ran into inconsistencies because he tried to combine the hypothesis of finite entropy with classical theory, but the hypothesis is viable only if one accepts the quantum-mechanical result that black holes emit thermal radiation" \cite{Hawking 1976}, p. 193.

\section{Hawking radiation} \label{radiation}

At about 1974, Hawking solved the above paradox by suggesting that black holes emit particles at a steady rate. The outgoing particles have a spectrum that is precisely thermal. The black hole creates and emits particles and radiation just as if it were an ordinary hot body with a temperature $T$ that is proportional to the surface gravity $\kappa$ and inversely proportional to the mass $M_{BH}$ [see equations (\ref{eq8}) and (\ref{eq9}) below] \cite{Hawking 1974}; \cite{Hawking 1977}, p. 37. This immediately solved all problems.

Hawking set $E = kT$, where $k$ is the Boltzmann constant. He showed that black holes emit precisely thermal radiation, like that of a black body, with temperature: 

\begin{equation} \label{eq8}
 T = \frac{\hbar \kappa}{2 \pi ck}.\end{equation}

From the point of view of an outside observer, a Schwarzschild black hole emits black body radiation with temperature $T$ that is proportional to $\kappa$ and inversely proportional to $M_{BH}$:

\begin{equation} \label{eq9}
 T = \frac{c^3 \hbar}{8 \pi GM_{BH} k} \approx \frac{6.2 \times 10^{-8} K}{m},\end{equation}
 
\noindent where $m$ stands for the black hole mass expressed in terms of solar mass $\approx 2 \times 10^{30} kg$:
 
\begin{equation} \label{eq10}
 m = \frac{M_{BH}}{M_\odot}.\end{equation}
 
Bekenstein said that to Hawking's surprise, by his own research he found that black holes are hot \cite{Bekenstein 2001}, p. 518. 

\section{The Bekenstein-Hawking entropy} \label{entropy}
 
Using equation (\ref{eq6}) Hawking wrote the black hole entropy as follows \cite{Hawking 1975}, p. 191, p. 197:

\begin{equation} \label{eq11}
 S_{BH} = \frac{k c^3}{G \hbar} \frac{A}{4},\end{equation}

\noindent where $S_{BH}$ is called the \emph{Bekenstein-Hawking entropy}. 

This equation is equivalent to Bekenstein’s equation (\ref{eq2}) or equation (\ref{eq24}). Hawking therefore fixed Bekenstein’s constant of proportionality $\eta = \frac{1}{4}$. 
The calculation, therefore, has been satisfactorily concluded and Wheeler alluded that Hawking ultimately came to the conclusion that Bekenstein was right after all \cite{Wheeler 2008}. 

In 1975 Hawking even wrote in his paper, "black hole thermodynamics": "A black hole of given mass, angular momentum, and charge can have a very large number of unobservable internal configurations which reflect the different possible configurations for the body that collapsed. If quantum effects were neglected, the number of different internal configurations would be infinite because one could form the black hole out of an indefinitely large number of indefinitely small mass particles. However, Bekenstein pointed out that the Compton wavelengths of these particles might have to be restricted to be less than the radius of the black hole and that therefore the number of possible internal configurations might be finite though very large" \cite{Hawking 1975}, p. 192. Recall that according to Thorne, most physicists (including Hawking and Thorne) had responded "Nonsense" to this idea of Bekenstein \cite{Thorne}, p. 426 (see Section \ref{Geroch}).

Delivering a lecture on his 60th birthday, Hawking outlined how he discovered Hawking radiation. At the end of 1973, Hawking's work with Penrose on the singularity theorems (\cite{Hawking and Penrose}, see Section \ref{hair}) had shown to him that general relativity broke down at singularities. So he thought that the obvious next step would be to combine general relativity, the theory of the very large, with quantum theory, the theory of the very small. He had no background in quantum theory so as a warm up exercise, he considered how particles and fields governed by quantum theory would behave near a black hole; he studied how quantum fields would scatter off black holes. "But to my great surprise", said Hawking, "I found there seemed to be emission from the black hole". At first he thought this must be a mistake in his calculation. But what persuaded him that it was real, was "that the emission was exactly what was required to identify the area of the horizon, with the entropy of the black hole [equation (\ref{eq11})].
\vspace{1mm} 

$S = \frac{Ak c^3}{4\hbar G}.$
\vspace{1mm} 

\noindent I would like this simple formula to be on my tombstone", said Hawking \cite{Hawking 2003}, pp. 112-113. 

In his 1975 paper, Hawking pointed out that Bekenstein's entropy $S_{BH}$ [\ref{eq2}] would have to be a function only of $M$, $J$, and $Q$ with the following two properties: first, it always increased when matter or radiation fell into the black hole. Second, when two black holes merged together, the entropy of the final black hole was bigger than the sum of the entropies of the initial black holes \cite{Hawking 1975}, p. 193. 
Bekenstein admitted that he very much guessed the form of the function in equation (\ref{eq2}): $S_{BH}=f(A)$ \cite{Bekenstein 2001}, p. 515. In 1975 Hawking remarked that "this is the only such quantity" \cite{Hawking 1975}, p. 193.  

I show in Section \ref{semi} that unlike Bekenstein who tried to unify thermodynamics and classical general relativity, Hawking endeavored to unify quantum mechanics and classical general relativity. And using a semi-classical approximation, Hawking arrived at equation (11).  

\section{The Bekenstein bound} \label{bound}

Bekenstein’s mind still boggled at the Geroch paradox and the GSL, see Section \ref{Geroch}. That is because even if black holes do carry entropy $S_{BH}$ and we add additional mass $dM$  to the black hole, the area of the horizon might not increase enough to compensate for the loss of entropy. The GSL could in this case be violated. 

Recall that the GSL is defined by equation (\ref{eq37}) and we lower down the box on the rope to the black hole and when it approaches $2R_g$, we release the radiation of high entropy $S$. But the GSL is violated $dS_g=dS_{BH}+S < 0$ because the black hole mass would not increase at all once the radiation from the box is finally dropped in $dM=0$ and the area of the black hole would not change $dS_M=S=0$. 

So, in 1981, Bekenstein set himself to solve this problem from a thermodynamic perspective. He set $G$, $c$, $k$ and $\hbar$ equal to 1. If a system of mass $M$, energy $E$, effective radius $R$ and entropy $S$ is dropped into a black hole, the mass increase of the black hole is: 

\begin{equation} \label{eq27}
dM_{BH} = \frac{ER}{4M_{BH}}. 
\end{equation}

This equation takes into consideration the gravitational redshift factor $\alpha$ given by equation (\ref{29}).

For a Schwarzschild black hole: $A=16 \pi M_{BH}^2$ [see equation (\ref{eq6})], so the increase in the surface area of the horizon of the black hole can be expressed as follows:

\begin{equation} \label{eq28}
dA=32 \pi M_{BH} dM_{BH}. 
\end{equation}

\noindent Inserting equation (\ref{eq27}) into equation (\ref{eq28}) one obtains: 

\begin{equation} \label{eq29}
dA=\frac{32 \pi M_{BH}ER}{4M_{BH}} = 8 \pi ER. 
\end{equation}

\noindent The GSL requires that $dS_g\geq 0$ and $dS_{BH} \geq -dS_M=S$. That is, according to equation (\ref{eq11}),$\frac{dA}{4} \geq S$.
\vspace{1mm} 

\noindent So, dividing equation (\ref{eq29}) by 4 yields:

\begin{equation} \label{eq30}
\frac{dA}{4}=\frac{8 \pi ER}{4},
\end{equation}

\noindent and the black hole entropy will increase by: $2 \pi ER$.
\vspace{1mm} 

Bekenstein stressed that if $S$ does not exceed $2 \pi ER$, the GSL is not violated. So, Bekenstein decided to impose a limit on $S$ saying that it cannot exceed what was later termed the Bekenstein bound: 

\begin{equation} \label{eq12}
S \leq 2 \pi ER.\end{equation}

\noindent $S_{BH}$ exactly saturates (satisfies) the Bekenstein bound \cite{Bekenstein 1974}, p. 3292; \cite{Bekenstein 1981}, p. 287; \cite{Bousso}, p. 145; \cite{Wald}, p. 7; \cite{Bain}.

\section{Buoyancy and Bekenstein's bound}

In response to Bekenstein's paper, William Unruh and Wald published a paper in which they showed that equation (\ref{eq12}) is not needed for the validity of the GSL \cite{Unruh and Wald}. 

Wald explains that Bekenstein’s interpretation of equation (\ref{eq2}) was that quantum effects prevent Geroch’s box from reaching $2R_g$. The quantum effects eventually turned out to be Hawking radiation. If a black hole is placed in a radiation bath of temperature smaller than that given by equation (\ref{eq8}), then Hawking radiation [with temperature given by equation (\ref{eq8}) or (\ref{eq9}) would dominate over accretion and absorption. And Bekenstein’s GSL would not be violated \cite{Wald}, p. 7.

More specifically, Unruh and Wald argue that if we attempt to slowly lower the Geroch box containing rest energy $E$ and entropy $S$ into a black hole, there will be an effective buoyancy force on the box caused by the acceleration radiation felt by the box when it is suspended near the black hole. As a result there is a finite lower bound on the energy delivered to the black hole in this process and a minimal area increase. This small area increase turns out to be just sufficient to ensure that the GSL is satisfied. Consequently, less work is delivered to infinity by the rope during the process of lowering the box and more energy is transferred to the black hole in this process than would occur classically. By the Archimedes principle of buoyancy this occurs when the energy of the box equals the energy of the displaced acceleration radiation \cite{Unruh and Wald}, p. 942, p. 944; \cite{Unruh2}, p. 155; see Section \ref{Geroch}.

Unruh and Wald wrote to Bekenstein to tell him about their finding. But Bekenstein had reservations about  the buoyancy effect and he gave examples where he showed that it was not sufficient to save the GSL. He stressed that his entropy bound on matter, equation (\ref{eq12}), would still be needed for the validity of the GSL. Unruh and Wald's solution seemed to have obviated the need for Bekenstein's bound. Indeed, Bekenstein remarked that Unruh and Wald's conclusion was that bound equation (\ref{eq12}) was not necessary for the GSL to hold. So, Bekenstein and Unruh and Wald had conflicting intuitions about the bound. Bekenstein, on his part, insisted that the fact remains that bound equation (\ref{eq12}) "is supported by direct statistical arguments". He further emphasized that "the bound must apply to any object that can be lowered to the horizon of a black hole" \cite{Wald}, p. 8; \cite{Bekenstein 1983}, p. 2262. \cite{Bekenstein 1994}, p. 1920. 

In 2002 Bekenstein explained that in the original derivation of equation (\ref{eq12}) he considered the Geroch process and then applied the GSL to get the bound. This derivation was criticized for not taking into account the buoyancy of the Geroch box. "A protracted controversy on this issue" led to the perception that a correction for buoyancy merely increases the $2 \pi$ coefficient in equation (\ref{eq12}) "by a tiny amount" \cite{Bekenstein 2002}, pp. 4-5. 

\section{Hawking’s semi-classical approximation} \label{semi}

Hawking discovered that black holes emit thermal radiation and gave a theoretical basis for the thermal radiation emitted by black holes: quantum effects cause the black hole to radiate. He used the semi-classical approximation to arrive at equation (\ref{eq8}). He wrote in 1978: "The conclusion that black holes radiate thermally was derived using a semiclassical approximation in which the matter fields are treated quantum mechanically on a classical space-time background" \cite{Hawking 1978}, p. 24. 

I shall start with Hawking's original derivation from 1974. In 1974 Hawking determined the number of particles created and emitted to infinity in a gravitational collapse. He considered a simple example in which the collapse of a star is spherically symmetric. Outside the field is described by the Schwarzschild solution. The asymptotically flat space-time is divided into three regions: 1) The future null infinity (light-like future infinity) $I^+$: all null geodesics can reach the future null infinity. 2) The horizon null surface of the black hole. 3) Past null infinity (light-like past infinity)$I^-$: no null geodesics can reach the future null infinity.   

Consider a wave which has a positive frequency on $I^+$ propagating backwards through space-time with nothing crossing the event horizon. Part of this wave will be scattered by the curvature of the static Schwarzschild metric outside the black hole and will end up on $I^-$ with the same frequency. Another part of the wave will propagate backwards into the star, through the origin and out again onto $I^-$. 

These waves can have a very large blue-shift and they can reach $I^-$ with asymptotic form of the wave function for time $v<v_0$ where $v_0$ stands for the last advanced time that enables them to leave $I^-$. These waves are then absorbed by the black hole. On the other hand, if $v>v_0$ the waves can pass through the origin and escape to $I^+$. These waves are red-shifted.  

Hawking calculated (took Fourier transforms of the asymptotic form) that the total number of outgoing particles or wave packets created in a certain frequency range (peaked at a certain frequency) is infinite. Hawking concluded that this infinite number of particles corresponds to a steady-rate emission, which means that the black hole will emit thermal radiation until it completely evaporates in an explosion (see the next section for black hole evaporation). 

Hawking compared the incoming waves that escaped to $I^+$ to the outgoing ones that arrived at $I^-$ and was led to the following equation, which represents the relation between absorption and emission (Hawking radiation):
\vspace{1mm} 

\noindent $ \left ( exp{\left ( \frac{2 \pi \omega}{\kappa} \right)}-1 \right)^{-1}, $
\vspace{2mm} 
\noindent that is, the number of particles emitted from the black hole in a wave packet mode times the number of particles that would have been absorbed from a similar wave packet incident on the black hole from $I^-$. 
"But this is just the relation", said Hawking, between absorption and emission that one would expect from a black hole emitting thermal radiation with a temperature given by equation (\ref{eq8}): $T=\frac{\kappa}{2 \pi}$. Hawking therefore justified the geometric reasoning made in black hole thermodynamics \cite{Hawking 1974}, pp. 30-31. 

In a paper he submitted in June 1975, "Black Holes and Thermodynamics", Hawking showed that the steady-rate emission turns out to have an exactly thermal spectrum and the following quantities arise from his 1974 semi-classical approximation: the area $A$ of the event horizon [equation (\ref{eq6})], the surface gravity $\kappa$ [equation (\ref{eq7})], the angular frequency of rotation of the black hole $\Omega$ and the potential of the event horizon $\Phi$ (both are inversely proportional to the area $A$ [see equation (\ref{eq3}) for the first law of black hole mechanics]). And the most important quantity that arises from the semi-classical approximation is the entropy of the black hole [equation (\ref{eq11})] \cite{Hawking 1975}, pp. 191-192.  

\section{Hawking’s ansatz of pair production}

Two months later, Hawking submitted a paper in which he was more explicit saying: "quantum mechanics allows particles to tunnel on spacelike or past-directed world lines. It is therefore possible for a particle to tunnel out of the black hole through the event horizon and escape to future infinity. One can interpret such a happening as being the spontaneous creation in the gravitational field of the black hole of a pair of particles, one with negative and one with positive energy with respect to infinity. The particle with negative energy would fall into the black hole [...]. The particles with positive energy can escape to infinity where they constitute the recently predicted thermal emission from black holes" \cite{Hawking 1976}, p. 2462. 

The paper was published in 1976 and a year later, Hawking explained that quantum field theory implies that particle-antiparticle virtual pairs (an electron and a positron) are continually produced and recombined in the vacuum state. Hawking showed that when such a particle-antiparticle pair is produced near the event horizon of a black hole, one particle may fall into the black hole, leaving the other particle outside without a partner with which to annihilate. “The forsaken particle or antiparticle may fall into the black hole after its partner”, says Hawking, “but it may also escape to infinity, where it appears to be radiation emitted by the black hole. 

Hawking wrote in 1978 that "If a black hole is present, one member of a pair may fall into the hole leaving the other member without a partner with whom to annihilate" \cite{Hawking 1978}, p. 23.

Another way of looking at this process, says Hawking, is to regard the particle that fell into the black hole (the antiparticle) as being a particle that was traveling backwards in time. It would come from the singularity and would travel backwards in time out of the black hole to the point where the particle-antiparticle pair first appeared. There it would be scattered by the gravitational field into a particle traveling forwards in time. Hawking reasoned that this way one can think of the radiation from a black hole as having come from the singularity and having quantum mechanically tunnelled out of the black hole \cite{Hawking 1978}, p. 32; \cite{Hawking 1977}, p. 37. In 1979 Hawking elaborated this picture by suggesting a radiation from white holes (see Section \ref{mini}).     

This way of looking at the pair-production process is reminiscent of the Richard Feynman-Ernst Stueckelberg interpretation of the positron, which is moving forward in time but appearing to us to be an electron traveling backward in time: a photon spontaneously creates an electron-positron pair, with an electron flying off to a distant region and the positron meeting an additional electron, resulting in mutual annihilation. This description involves three particles: electron, positron, and another electron. Feynman suggested describing that same process using just one electron moving first forward in time, then backward in time and finally again forward in time \cite{Feynman} p. 749.

Hawking concludes that quantum mechanics allows a particle to escape from inside a black hole, something that is not allowed in classical mechanics \cite{Hawking 1977}, p. 38. Simply put, from the point of view of an observer at infinity: the black hole creates particles in pairs, with one particle always falling into the black hole and the other escaping to infinity \cite{Hawking 1976}, p. 2460.

This pair production picture rings a familiar bell. Recall that Penrose and Floyd had suggested a process in which a particle with four-momentum is dropped into the ergosphere. It then splits into two particles, so that the mass-energy of the first particle is negative and it is swallowed by the black hole whilst the second particle escapes back to infinity and carries more mass-energy to infinity. Penrose and Floyd showed that the area theorem is valid. See Section \ref{area}. On the other hand, as I show in the next section, Hawking radiation violates the area theorem. Both processes, however, do not violate the GSL. More on the GSL in Section \ref{GSL2}. 

Unruh has recently said that one of the problems with Hawking's derivation is that although it is mathematically correct, it really makes no physical sense. There is nothing wrong with the mathematics but physically it simply makes no sense. Why is it problematic? The particles (Hawking radiation) cannot come within the black hole because nothing can come out of the black hole. They are actually created outside of the black hole but an observer at infinity sees the black hole as if it is emitting a Hawking radiation \cite{Unruh3}.     

\section{Hawking’s evaporation time} \label{evaporate}

Now let us obtain the power $P$ emitted by a Schwarzschild black hole of mass $M_{BH}$. But when obtaining $P$ there is a snag. As Hawking himself put it in 1974: as the black hole emits the thermal radiation with temperature $ \approx 10^{-6}$ [see equations (\ref{eq8}) and (\ref{eq9})] “one would expect it to lose mass. This in turn would increase the surface gravity and so increase the rate of emission. The black hole would therefore have a finite life of the order of $10^{71} \left (\frac{M \odot}{M} \right )^{-3}s$. 
\vspace{1mm} 
For a black hole of solar mass this is much longer than the age of the Universe” \cite{Hawking 1974}, p. 30.
\vspace{1mm} 

The Stefan-Boltzmann law of black body radiation states that the radiated power $P$ per unit area of a black body is given by: 

\begin{equation} \label{eq31}
P = \sigma \epsilon A T^4, 
\end{equation}

\noindent where $\sigma T^4$ represents the flux at which a black body emits, $\epsilon =1$ because a black hole is a perfect black body, and $\sigma$ stands for the Stefan-Boltzmann constant: 

\begin{equation} \label{eq32}
\sigma = \frac{\pi^2 k^4}{60c^4\hbar^3}.
\end{equation}
\vspace{1mm} 

We combine equations (\ref{eq31}) and (\ref{eq32}) and equations (\ref{eq6}) and (\ref{eq9}) to obtain the rate at which a black hole emits energy across a surface with an area of the black hole horizon, from the point of view of an outside observer:

\begin{equation} \label{eq13}
P = -\frac{dE}{dt}=\frac{\hbar c^6}{15360 \pi G^2 {M_{BH}}^2} = \frac{9 \times 10^{-22}}{m^2}\frac{erg}{s},\end{equation}

\noindent where $m$ is given by equation (\ref{eq10}). This is terribly low.
Inserting:

\begin{equation} \label{eq14}
E = M_{BH} c^2,\end{equation}

\noindent into equation (\ref{eq13}), the left-hand side can be written as follows:

\begin{equation} \label{eq33}
-\frac{dE}{dT}=-c^2\frac{dM_{BH}}{dt}.     
\end{equation}

\noindent Rearranging the terms in equation (\ref{eq13}) yields:

\begin{equation} \label{eq34}
-\frac{dM_{BH}}{dt}=\frac{\hbar c^4}{15360 \pi G^2 {M_{BH}}^2}, 
\end{equation}

\noindent from which we obtain the evaporation time $t$ for the black hole, from the point of view of an outside observer:

\begin{equation} \label{eq15}
t= \frac{5120\pi G^2 {M_{BH}}^3}{\hbar c^4}\approx m^3\times 6.67 \times 10^{74} s,\end{equation}

\noindent where $m$ is given by equation (\ref{eq10}). This is a tremendous amount of time. 

When particles escape from the black hole, the black hole loses a small amount of its energy and consequently, according to equation (\ref{eq14}), its mass decreases. Its temperature $T$ increases as a result of emitting Hawking radiation. It therefore radiates more Hawking radiation, loses mass $M_{BH}$, becomes again hotter and radiates even faster until it finally evaporates. The smaller the black hole gets, the higher its temperature $T$ goes and in turn the loss of mass $M_{BH}$ accelerates. By the time the black hole reaches the size of a micro black hole with a Planck mass:   

\begin{equation} \label{eq16}
m_p=\sqrt{\frac{\hbar c}{G}}=2.176\times 10^{-8} kg,
\end{equation}

\noindent it almost evaporates completely and its temperature $T$ is so high that it will explode. Which means that Hawking’s area theorem (the second law of black hole mechanics) is violated because the area of the black hole gets smaller and smaller as the black hole radiates Hawking radiation.

\section{Saving the GSL from refutation} \label{GSL2}

Hawking radiation unveiled an unexpected possibility. The black hole is emitting Hawking radiation, is losing mass and entropy in the course of time. Consequently, Hawking radiation (and evaporation) slowly drains the black hole mass-energy and the black hole entropy decreases. That is because according to equation (\ref{eq11}), $S_{BH} \propto {M_{BH}}^2$. This posed a problem for the GSL.

In 1975 Hawking set himself to solve this problem. First, given the entropy of a system as a function of the energy $E$ of the system, the temperature can be defined as:

\begin{equation} \label{eq17}
\frac{1}{T}= \frac{\partial S}{\partial E}.
\end{equation}

\noindent Thus, the temperature of a black hole can be defined as:

\begin{equation} \label{eq18}
\frac{1}{T_h}= \frac{\partial S_h}{\partial M_{B_H}},
\end{equation}

\noindent where $S_h$ is the entropy of the black hole and $T_h$ is the temperature of the black hole.

Second, says Hawking, the GSL is then equivalent to the requirement that heat should not run from a cooler system to a warmer one.   

So, Hawking returned to the thought experiment of a black hole which is surrounded by black body radiation at some temperature $T_M$ (see Section \ref{bath}). For any nonzero $T_M$ there will be some rate of accretion of this radiation into the black hole. 

Now suppose the black hole is in a hotter radiation bath, $T_M>T_h$, it follows from equation (\ref{eq18}) that $dS_M<dS_h$. $dS_M$ is caused by the accreting radiation. In this case, the GSL holds. The black hole can absorb matter but does not emit anything. Hawking remarked that this case holds in purely classical theory \cite{Hawking 1975}, p. 193. 

On the other hand, if the black hole is in a colder radiation bath, $T_M<T_h$, the GSL is violated. In classical theory, one sets $T_h=0$ in equation (\ref{eq18}), and so $S_h$ is infinite. In this case, equation (\ref{eq2}) is meaningless. One inevitably runs into inconsistencies. Hawking concluded that the black hole has to emit thermal radiation with some finite nonzero temperature. 
 
So, if $T_M>T_h$, the accretion onto the black hole is greater than the Hawking radiation and $dS_h>dS_M$. But if $T_M<T_h$, the emission of Hawking radiation is greater than the accretion onto the black hole and $dS_M>dS_h$. Hawking then explained that the fact that the black hole emits Hawking radiation with a temperature [equation (\ref{eq8})] enables one to establish the GSL and prove that the entropy $S_h$ [equation (\ref{eq11})] is finite. Hawking integrated equation (\ref{eq8}). This allows one to integrate the first law of black hole mechanics, equation (\ref{eq4}) and deduce that:

\begin{equation} \label{eq35}
S_h=\frac{1}{4} A + const.     
\end{equation}

If one makes the reasonable assumption that the entropy tends to zero as $M_{BH} \rightarrow 0$, the constant in equation (\ref{eq35}) must be zero. Thus, the fact that the black hole emits quantum radiation with a temperature given by equation (\ref{eq8}), enables one to prove the GSL and establish that $S_h$, equation (\ref{eq11}), is finite \cite{Hawking 1975}, p. 193.

In 1975 Bekenstein reanalyzed the same thought experiment of the black hole immersed in a black body radiation bath within a cavity of temperature $T_M$, sufficiently low in relation to $T_{BH}$. $S_M$ is the entropy of radiation and accreting matter outside the black hole. Bekenstein reasoned that the flow of radiation from the cavity into the black hole (lower to higher temperature) will violate the GSL unless some process generates entropy outside the black hole. In 1974 he had suggested a cumbersome and illogical mechanism to accomplish this, see Section \ref{bath}. But in 1975 he recognized that Hawking radiation was a simpler resolution of the difficulty. So, Bekenstein showed that the GSL in averaged form is respected: after the radiation emitted by the black hole is assimilated into the ambient bath, $S_{BH}$ plus $S_M$ will be larger than before. Although Hawking radiation causes the surface area $A$ to steadily decrease, eventually the mean $dSg=dS_{BH}+dS_M>0$ [equation (\ref{eq37})] for all $T$ and vanishes only when $T_M=T_{BH}$ \cite{Bekenstein 1975}, p. 3079. 

Bekenstein realized that the only way the GSL could hold was for the Hawking radiation to carry enough entropy to overcompensate for the loss \cite{Bekenstein 2001}, p. 518. During the radiation process, the area of the Kerr black hole can decrease but the radiation entropy increases [see equations (\ref{eq 25}) and (\ref{eq 26}) for the area of a Kerr black hole]. Bekenstein gives the analogy between a black hole and a hot body: just as for the hot body, one can expect that the black hole entropy plus the radiation entropy would most likely increase \cite{Bekenstein 1975}, pp. 3077-3078.

Bekenstein explained that the GSL was designed to replace the ordinary second law of thermodynamics. Accordingly, the GSL makes a stronger statement than Hawking's area theorem. The area theorem requires only that $A$ not decrease (see Section \ref{GSL}). By contrast the GSL, "being an intrinsically quantum law could be expected to fare better. And indeed in the astonishing quantum process of spontaneous radiation by a Kerr black hole discovered by Hawking, the area theorem is flagrantly violated, but the increase in exterior entropy due to the radiation is expected to suffice to uphold the GSL". The GSL is not exactly "an intrinsically quantum law". But Bekenstein considered Hawking's proof as a "confirmation of the validity of the GSL for a process not even dreamt of at the time of its inception" and stressed that this is a "striking evidence of the versatility" of the GSL and of "the physical meaningfulness of the concept of black-hole entropy which underpins it" \cite{Bekenstein 1975}, p. 3077.

In his 1975 paper, Hawking did not forget Bekenstein when he wrote that the latter ran into inconsistencies because he tried to combine the hypothesis of finite entropy with classical theory \cite{Hawking 1975}, p. 193. Bekenstein though understood things quite differently. He stressed that his argument (from 1975) is implicit in Hawking's suggestion that the GSL is respected in the black hole radiation process. Of course, Bekenstein emphasized that it was Hawking who "gave a simple proof" that $dSg=dS_{BH}+dS_M$ "is larger than before", i.e. that equation (\ref{eq37}) holds. In other words, Hawking gave a resolution to the difficulty posed by the case $T_M<T_{BH}$ and by that salvaged the GSL from refutation (see Section \ref{entropy}). Bekenstein was pleased as ever. He concluded that Hawking has "started as a vocal critic of" the GSL but "became the person who made it fully consistent with" the thought experiments \cite{Bekenstein 1980}, p. 29.

\section{Mini black holes and experimental verification} \label{mini}

Hawking radiation is widely accepted but we still lack observational evidence confirming Hawking’s predictions. According to equation (\ref{eq9}), the temperature of Hawking radiation for a stellar mass black hole is formidably law and about six to eight orders of magnitude below that of the cosmic microwave background radiation. This means that we cannot detect Hawking radiation from stellar-mass black holes, let alone from supermassive black holes. Put differently, as Hawking himself was well aware in 1974, according to equation (\ref{eq15}), the evaporation time is proportional to ${M_{BH}}^3$ and a black hole of one solar mass $M_\odot\approx 2 \times 10^{30}g$ will have evaporation time of more than $t \approx 2\times 10^{67}y$, see Section \ref{evaporate}. This is much longer than the age of the universe $\approx 10^{10}y$. The minimum mass of a stellar-mass black hole is $3.3 M_\odot$ and for the supermassive black hole at the center of our galaxy the evaporation time is approximately $10^{94}y$ . The evaporation time is in fact longer because black holes are absorbing matter whilst gradually shrinking to awfully small sizes. 

So, maybe we can detect Hawking radiation from micro black holes? A black hole mass cannot be below a Planck mass, see equation (\ref{eq16}). That is because if it were, the black hole would be smaller than its own Compton length:

\begin{equation} \label{eq36}
\lambda_c = \frac{\hbar}{m_p c},     
\end{equation}

\noindent which is equal to its Schwarzschild radius $2l_p$ [$2R_g$, see equation (\ref{eq23})] and is of the same order as $l_p$: 

\begin{equation} \label{eq19}
l_p=\sqrt{\frac{G\hbar}{c^3}}\approx 1.616 \times 10^{-35}m.
\end{equation}

\noindent Consequently, the black hole would not exhibit the black hole hallmark, the event horizon \cite{Bekenstein 2004}, p. 32.

\noindent The Planck time is: 

\begin{equation} \label{eq20}
\frac{G^2{m_p}^3}{\hbar c^4}=t_p=\sqrt{\frac{G\hbar}{c^5}}\approx 5.39 \times 10^{-44}s,\end{equation}

\noindent and according to equation (\ref{eq15}), the evaporation time $t$ for a Planck-size black hole is:

\begin{equation} \label{eq21}
t=\frac{5120\pi G^2 {m_p}^3}{\hbar c^4}=5120\pi t_p\approx 8 \times 10^{-40} s.\end{equation}

So, an extremely small black hole radiates all its mass-energy in a tiny amount of time. 

In 1974 Hawking wondered whether there might "be much smaller black holes which were formed by fluctuation in the early Universe. Any such black hole of mass less than $10^{15} g$ would have evaporated by now. Near the end of its life the rate of emission would be very high” (see explanation in Section \ref{evaporate}). Hawking conjectured that in the last $0.1s$ the black hole’s rate of emission would be $10^{30} erg$. Although by astronomical standards, says Hawking, the explosion of this black hole would be fairly small, it is equivalent to about a million hydrogen bombs of one megaton each  \cite{Hawking 1974}, p. 30.

Hawking therefore suggests that we detect Hawking radiation from hypothetical primordial mini black holes (about the size of a mountain on Earth) with $M_{BH}\leq 10^{12} g$. A hypothetical $10^{12} g$ primordial black hole, which might have been formed in the early universe, would radiate its mass over the age of the universe and would now be entering its final stages of evaporation. The evaporation time for such a low-mass black hole is $t \approx 2.267 \times 10^9 y$. If such tiny black holes in fact exist, then we could see them evaporate in an explosion.

In 1975 Hawking suggested that the mini black holes are equivalent to the hypothetical white holes. The time reverse of black holes, says Hawking, a white hole, must also occur. While black holes emit Hawking radiation and their event horizons swallow matter and radiation, white holes emit matter and radiation and nothing can enter into their event horizons. Hawking explains that white holes emit radiation at the same rate as black holes. That is, "if one makes the reasonable assumption that the emission of a white hole is independent of its surroundings, it follows that it is the same as the rate of emission of a black hole of the same mass, angular momentum, and charge". Hawking concluds that black and white holes are identical to external observers. In 1979 Hawking said in an interview: "I think there's only one entity. There are only holes. They appear black when they are big and white when they are small" \cite{Overbye}, p. 107; \cite{Hawking 1975}, p. 196. 

\section{Unruh's acoustic black holes}

The inability to detect Hawking radiation led Unruh in 1981 to suggest that one could detect Hawking radiation in the laboratory. Unruh proposed the following idea: let us create a small artificial black hole in the laboratory whose physics and mathematics is analogous to that of the astrophysical black hole. The artificial black hole is based on the idea of sound waves propagating in a moving fluid. The waves are dragged by the fluid. Co-propagating waves move faster and counter-propagating waves are slowed down. At some point the fluid exceeds the speed of the waves and beyond this point, no counter-propagating wave would be able to propagate upstream. The point where the moving fluid matches the wave velocity is an event horizon \cite{Unruh}, p. 1351, p. 1353; \cite{Leonhardt}, pp. 2852-2853. 

But the acoustic black hole is hobbled by a serious problem. It cannot reproduce $S_{BH}$, the Bekenstein-Hawking entropy, equation (\ref{eq11}). The reason is that Hawking radiation is a purely kinematic effect that depends only on the existence of a certain metric and some sort of horizon. In contrast, $S_{BH}$ is associated with dynamical effects, i.e. the area of the event horizon. According to the first law of black hole mechanics, entropy equals one quarter the area only if the Einstein field equations are valid \cite{Visser}, p. 651.

\section{Penrose's Hawking points}

Penrose has raised the following suggestion. We know that the Hawking temperature of black holes of several solar masses is extremely cold and the Hawking radiation they emit is ridiculously tiny. We will therefore not see any significant effect in these black holes. Penrose has asked: how can we measure or detect Hawking radiation? He suggests that we have to wait until the universe expands and expands and the ambient temperature goes down and down and ultimately gets less than the temperature of the black hole. Then the black hole starts to lose its energy and, therefore, it loses its mass and ultimately it would evaporate. When will this happen? Penrose answers that this will probably happen in years. Of course, this seems like an awfully long time for stellar-mass black holes and Penrose gathers that with regards to supermassive black holes that we now see, they have a much longer time than that must be expected. But he proposes a cyclic picture in which we actually do not need to wait this immensely long time because we can observe the Hawking radiation in the form of what he calls \emph{Hawking points} in our present universe. Let us Pause for a moment on Penrose’s hypothesis.

Penrose claims that if we are waiting far way-way into the future, as the universe expands, clusters of galaxies tend to cling together, and every now and again the supermassive black holes will run into each other and merge. For instance, our galaxy is on a collision course with the Andromeda galaxy. It will not be for three thousand million years from now so do not worry, says Penrose. But when the galaxies will eventually collide, the black holes will spiral around into each other and there will be a great explosion ending up with a huge black hole, much bigger than the one at the center of the Andromeda galaxy.

The majority of material in the galaxies will ultimately be swallowed by the huge black holes sitting around the clusters of galaxies and gravitational waves will carry the energy. As the universe expands and expands, in the very-very far future, in googolplex years or so, says Penrose, what will eventually remain are big black holes that will ultimately disappear. They will disappear with what Penrose calls a “pop” (an explosion) and almost the entire mass-energy of the black holes goes in the form of Hawking radiation of photons.

Now, this is a thing, says Penrose, that used to worry him because what we have got is a universe full of photons. On the whole as far as particles numbers are concerned, in the remote future, the universe would completely be dominated by cold photons, which are conformally invariant. Ultimately the mass asymptotically goes to zero.

In the big bang, on the other hand, you have the opposite situation because we have very high temperatures. When those temperatures get higher and higher, the energy of individual particles is completely dominated by the motion and the mass becomes utterly unimportant. Hence, in this limit again, you would have massless things. 

Penrose put forth the following hypothesis called the conformal picture: as you approach the boundary at both ends – the big bang and eternity, the argument is that the cosmic conformal cyclic cosmology allows a crossover from one “aeon” (universe) to the next, to be described in terms of the conformal physics. Penrose argues that we in our universe are looking here through the big bang of the previous universe, aeon, which is due to Einstein’s cosmological constant.

How can we confirm that very hypothesis of Penrose? Penrose suggests that his hypothesis may be confirmed by the Hawking points. Before explaining what are these points, Penrose says there are lots of mathematical details which need to be worked out and he has not yet done so. Hence, his explanation is very general and his theory is not fully baked.

Penrose explains that rather than thinking about going into the future, we think about going into the past, the big bang. We know that according to the second law of thermodynamics, the entropy increases in time. The same statement of the second law says that as we go back and back in time into the past, the entropy is supposed to go down and down. The best evidence for the big bang that we have is the microwave cosmic background radiation which comes from all directions. What we see at this early stage, says Penrose, is maximum entropy and this is quite evident as the main feature of this radiation and its spectrum of black body radiation. So, Penrose says he thought there was something fishy about that because at the early universe, as far as gravity is concerned, the universe expanded and expanded, matter clumped into galaxies and black holes were created and black holes represent maximum entropy. On the other hand, the universe started with very low entropy at the beginning.

Penrose’s conjecture consists of the following main idea: we have a huge explosion of the supermassive black hole, one burst which causes the evaporation of the supermassive black hole from the previous universe, aeon. The evaporation basically leads to photons that spread out. But where can we find these photons? When you think of how far does this radiation spread out, says Penrose, the answer is, almost nothing. So, all the Hawking radiation from the evaporation is eventually concentrated into one point. This entire Hawking radiation over years is simply squashed down into one tiny little point, a Hawking point, probably smaller than the Planck scale. The Hawking points are created just when inflation is turned off, as we go back again to the big bang. Accordingly, we have a burst of radiation coming through at that little point. So, it is a matter of a cyclic model of stretching and squashing, stretching and squashing of aeons.

Hawking once argued that the black holes eventually evaporate away and information must be lost in the process, see Sections \ref{demon2} and \ref{loss}. Though over the years he backed off this claim, Penrose has adopted it. According to the principle of unitarity of quantum mechanics, you cannot lose information. But, says Penrose, unitarity is violated by the way we use quantum mechanics when we make measurements, see Section \ref{loss}. 
Penrose says he can make very cogent reasons why the information is lost and this is related to his hypothesis because when black holes disappear the effective entropy comes down and it means that we arrive at the beginning of the universe, the big bang.

Penrose admitted that he would often “talk about these things through lectures and I thought well, it’s perfectly safe, I could talk about conformal cyclic cosmology until the end of time because nobody will ever be able to disprove me. Then I began to think, is that true?” Maybe we can have a test of conformal cyclic cosmology? Maybe one can detect Hawking radiation in the form of spots that came about and spread from the big bang and three hundred and eighty thousand years afterwards? It means that what Penrose calls the Hawking points, the Hawking radiation concentrated in those spots at 4 degrees Kelvin, spread out from the big bang to a much more uniform distribution to what we actually see in the microwave cosmic background radiation. If you happen to ask Penrose: “Why haven’t people seen these before?” His answer is plain and simple: “Because they haven’t been looking!” \cite{Penrose3}, \cite{Penrose4}, \cite{Penrose5} and \cite{Penrose6}.

Suppose we detected hot (4K) spots in the cosmic microwave background radiation. We cannot say that those spots are a conclusive evidence for Penrose’s cyclic universe, namely, clear-cut evidence for Hawking radiation from a universe before the big bang. Those spots could be explained by as many as possible models. This means that one major evidence cannot support an entire theory.

\section{Evaporation and the censorship hypothesis}

In 1969 Penrose pondered whether there exists a “cosmic censor” who forbids the appearance of naked singularities, clothing each one in an event horizon. Penrose answered that it is not known whether naked singularities would ever arise in a collapse which starts off from a non-singular initial state \cite{Penrose}, p. 1160. 

In 1975 Hawking formulated \emph{“The ‘cosmic censorship’ hypothesis: Nature abhors a naked singularity”}, namely, any singularities which are developed from gravitational collapse will be hidden from the view of observers at infinity by an event horizon. Hawking pointed out that evaporation violates the classical censorship hypothesis. He explained that if one tries to describe the process of a black hole losing mass and eventually disappearing and evaporating by a classical space-time metric, then “there is inevitably a naked singularity when the black hole disappears. Even if the black hole does not evaporate completely one can regard the emitted particles as having come from the singularity inside the black hole and having tunneled out through the event horizon on spacelike trajectories” \cite{Hawking 1976}, p. 2461. 

So, Penrose said that if “the singularity is visible, in all its nakedness, to the outside world!” \cite{Penrose}, p. 1160, God forbid, Hawking answered that “an observer at infinity cannot avoid seeing what happens at the singularity” \cite{Hawking 1976}, p. 2461.

\section{Bekenstein summons Maxwell's demon} \label{demon2}

According to the classical no-hair theorem and general relativity, information that enters the black hole is lost forever. Since the black hole evaporates, $\hbar \rightarrow 0$. According to equations (\ref{eq8}) and (\ref{eq18}), $T_{BH}\rightarrow 0$ and $S_{BH}\rightarrow 0$. At the end there is only Hawking radiation with no black hole. The question we need to be asking is: what happens to the information about the actual internal configuration of the back hole?

Hawking thus had been wracking his brain about a new problem that popped out: a black hole of $M_{BH}$, $J$, and $Q$ can have a large number of different unobservable internal configurations which reflect the possible different initial configurations of the matter which collapsed to produce the black hole. The logarithm of this number can be regarded as $S_{BH}$. $S_{BH}$ is a measure of the amount of information about the initial state which was lost in the formation of the black hole  \cite{Hawking 1975}, pp. 191. 

In fact Bekenstein was the first to recognize the relationship between information and black hole entropy. Unlike Hawking, Bekenstein defined entropy as follows (see Section \ref{entropy2}): $S_{BH}$ can be regarded as our ignorance or lack of information about the actual internal configuration of the back hole.  

Bekenstein was troubled by Hawking radiation, which seemed to him to have led to a deep problem. A black hole is prepared from matter in pure state but then it evaporates and radiates away its mass in the form of thermal Hawking radiation. Bekenstein found two contradictions: 

\emph{The first contradiction}: One is left with a high-entropy mixed state of Hawking radiation (see Section \ref{GSL2}). Bekenstein said that he found a contradiction in Hawking's attempt to provide an interpretation of the black hole entropy as the measure of the information hidden in the black hole, i.e. the information about the ways the black hole might have been formed. But this was Bekenstein's interpretation, not Hawking's! Bekenstein went on to say that the final state of the black hole has a lot of entropy and therefore we are faced with a large intrinsic loss of information. The way Bekenstein saw it was that a thermal radiation is incapable of conveying detailed information about its source. Consequently, hidden information remains sequestered as the black hole radiates, and when the black hole finally evaporates away, the information must be lost forever \cite{Bekenstein 1993}, p. 3680.  

In order to resolve this paradox Bekenstein showed that it would be necessary to summon Maxwell's demon. Bekenstein first exorcised Wheeler's demon (see Section \ref{demon}) and then he summoned Maxwell's demon. This was the second time that Bekenstein raised Maxwell's demon (see Section \ref{entropy2}).

Leo Szilard wrote in 1929: "A measurement procedure underlies the entropy decrease effected by the intervention of intelligent beings", \cite{Szilard}, p. 304.  
In 1993 Bekenstein explained that "just as in Szilard's famous discussion of Maxwell's demon where acquisition of information about the location of the molecule in the box was tantamount to the 'gas' having less entropy than expected, so here, the gradual information outflux is tantamount to the black hole entropy becoming gradually less and less than originally expected". Bekenstein concluded that indeed, the black hole entropy [equation (\ref{eq11})] decreases. But Hawking proved in 1975 that this was not the case (see Section \ref{GSL2}).

In 1976, Hawking suggested “a quantum version of the ‘no hair’ theorems” which “implies that an observer at infinity cannot predict the internal state of the black hole apart from its mass, angular momentum, and charge: If the black hole emitted some configuration of particles with greater probability than others, the observer would have some a priori information about the internal state” \cite{Hawking 1976}, p. 2462. In 1978 Hawking argued that a large amount of information is irretrievably lost in the formation of a black hole. The number of bits of information lost can be identified with the entropy of the black hole. Because part of the information about the quantum state of the system is lost down the black hole, the final situation after the black hole has evaporated is described by a mixed state rather than a pure quantum state \cite{Hawking 1978}, pp. 24. 
 
\section{Black hole information loss and noise} \label{loss}

Szilard wrote: "A measurement procedure". Does this hold for black holes? Bekenstein was led to: \emph{The second contradiction}: It seemed to Bekenstein that "a pure state could be converted into a mixed one through the catalysing influence of a black hole!" \cite{Bekenstein 2004}, p. 35. This contradicts the quantum principle of unitarity that says that a pure state will always remain pure under unitary evolution. "Hawking was led by this conclusion to assert that gravity violates the unitarity principle of quantum theory". "To be sure", said Bekenstein, "Hawking’s inference has remained controversial: whereas general relativity investigators have tended to accept this conclusion, particle physicists have stood by the unitarity principle and orthodox quantum theory" \cite{Bekenstein 2004}, p. 35.

Let us leave aside the mathematics for the moment. According to quantum mechanics, a pure quantum state is a solution of the deterministic time-dependent Schrödinger equation. The complete information of a state at one point of time allows us to determine the state at all other times. And so, the unitary evolution is reversible. Hence, no information is lost and we must be able to invert things and retrieve all information that went into the black hole, so long as no measurement is made on the system. But that is exactly the point; or maybe not. The thing is that the initial pure state that formed the black hole cannot be reconstructed. That is because the pure state has evolved into a mixed state. And we measure a mixed state. But what if we did not exist? Carlo Rovelli pointed out that according to Shannon's definition of information (the one adopted by Bekenstein, see Section \ref{entropy2}), "We do not need a human being, a cat, or a computer to make use of this notion of information" \cite{Rovelli}, p. 1641. Dozens of papers have been published in scientific journals on the information loss paradox and possible solutions to this paradox.  

Back to Bekenstein. "The two contradictions", concluded Bekenstein, "are facets of the black hole information loss paradox" \cite{Bekenstein 1993}, p. 3680.

From Hawking's perspective, the information paradox concerned the following contradiction: “This is the information paradox: How does the information of the quantum state of the infalling particles re-emerge in the outgoing radiation? This has been an outstanding problem in theoretical physics for the last forty years” \cite{Hawking 2015} p. 2. 

In 2015 Hawking tried to propose the following solution: “The information about the ingoing particles is returned, but in a highly scrambled, chaotic and useless form. This resolves the information paradox. For all practical purposes, however, the information is lost” \cite{Hawking 2015}, p. 3. Hawking had already suggested this solution to the information loss paradox in 1978: "The loss of information is equivalent to
the acquisition of new random information or 'noise'"\cite{Hawking 1978}. p. 24.

\section{Discussion: Bekenstein and Hawking}

In response to Wheeler's tea cups thought experiment, Bekenstein suggested on thermodynamic grounds that a black hole should have a finite entropy that is proportional to the surface area of the event horizon. Bekenstein formulated the GSL and inaugurated the field of black hole thermodynamics.  

However, Bekenstein then discovered a problem. Geroch invented a thought experiment which could violate Bekenstein’s GSL. Suppose a massless box is filled with radiation and then slowly lowered by a massless rope toward the horizon of a Schwarzschild black hole in its gravitational field. It is opened at $r\rightarrow 2R_g$ and the radiation is allowed to escape into the black hole. The amount of work that can be extracted from the Geroch heat engine is $Mc^2$ and the black hole would end up in the same state in which it began and all of the energy of the radiation could be converted to work in the laboratory from which one was doing the lowering. If all the mass-energy is converted to work, then black holes are systems at absolute zero temperature. In an attempt to salvage his GSL, Bekenstein assumed that the black hole entropy would be enormous and he gave a contrived quantum explanation for why the box would never reach $2R_g$. 

This argument seemed implausible to Hawking because if the effective temperature of a black hole is zero, then this means that one can in principle add entropy to a black hole without changing it in any way. Bekenstein responded to this claim by assigning to the black hole a finite non-zero temperature that corresponds to the finite black hole entropy. Bekenstein's temperature of the black hole is proportional to the surface gravity. 

Hawking on his part maintained that this in turn indicated that a black hole could be in equilibrium with thermal radiation at some temperature other than zero. But the problem is that black holes which can devour matter do not emit anything. So Bekenstein tried to solve this paradox in terms of his GSL and the second thought experiment was invented. 

Imagine a black hole which is surrounded by thermal radiation at some temperature. If the black hole is in a hotter radiation bath, the GSL holds. The black hole can absorb matter but does not emit anything. On the other hand, if the black hole is in a colder radiation bath, the GSL is violated. Bekenstein gave a cumbersome explanation for why his GSL was not violated in this case. Hawking thought that this resolution was far-fetched because no equilibrium of a black hole with thermal radiation at some temperature other than zero was possible. The black hole would absorb any thermal radiation that fell onto it but would be unable to emit anything in return. 

Hawking was now faced with the following problem: either the temperature of the black hole is identically zero, in which case its entropy is infinite and the concept of black hole entropy is meaningless, or black holes have to emit thermal radiation with some finite nonzero temperature. Hawking realized that the first case is what holds in purely classical theory, in which black holes can absorb but do not emit anything. Hawking argued that Bekenstein ran into inconsistencies because he tried to combine the hypothesis of finite entropy with classical theory. 

Accordingly, in 1974 Hawking offered the following resolution: the paradoxes are solved only if one accepts the quantum mechanical result that black holes emit thermal radiation. Hawking gave an explanation for "how the thermal radiation arises". He used the semi-classical approximation to arrive at equation (\ref{eq8}). That is, he combined quantum field theory and classical general relativity and obtained equation (\ref{eq8}) \cite{Hawking 1974}, p. 30. 

In June 1975, Hawking showed that the steady-rate emission of Hawking radiation turns out to have an exactly thermal spectrum and the following quantities arise from his 1974 semi-classical approximation: the area of the event horizon, the surface gravity of the black hole, the angular frequency of rotation of the black hole and the potential of the event horizon (both are inversely proportional to the the area, as expressed by the first law of black hole mechanics). And the most important quantity that arises from the semi-classical approximation is the entropy of the black hole \cite{Hawking 1975}, pp. 191-192. Hawking therefore used the semi-classical approximation to obtain black hole thermodynamic quantities.   

In August 1975, Hawking submitted a paper in which he gave a thorough theoretical explanation for Hawking radiation based on the preliminary explanation he had given in 1974 \cite{Hawking 1976}. 

Underlying this story is a sense of dispute between two opposing worldviews: that of Bekenstein guided by black hole thermodynamics, and that of Hawking, led by quantum field theory.   

In 1974 Hawking wrote: “Bekenstein suggested on thermodynamic ground that some multiple of $\kappa$ should be regarded as the temperature of the black hole. He did not, however, suggest that a black hole could emit particles as well as absorb them” \cite{Hawking 1974}, p. 31. When Hawking did the calculation, as he wrote in his well-known book, \emph{A Brief History of Time}, “I found, to my surprise and annoyance, that even nonrotating black holes should apparently create and emit particles at a steady state. At first I thought that this emission indicated that one of the [semi-classical] approximations I had used was not valid. I was afraid that if Bekenstein found out about it, he would use it as a further argument to support his ideas about the entropy of black holes, which I still did not like. However, the more I thought about it, the more it seemed that the approximations really ought to hold” \cite{Hawking 1988} p. 105.

Bekenstein, on his part, maintained that "Hawking had been a leader of the vociferous opposition to black hole thermodynamics". He believed that Hawking was trying to discredit his black hole thermodynamics. Bekenstein said that in his paper with Bardeen and Carter, Hawking "argued \emph{against}" Bekenstein's entropy and "Ironically, many uninformed authors still cite" the  Hawking, Bardeen and Carter paper "as one of the sources of black hole thermodynamics!" \cite{Bekenstein 2001}, p. 518. 

Bekenstein praised Hawking's achievement but said that Hawking “provided the missing pieces of black hole thermodynamics” \cite{Bekenstein 1980}, p. 29. Which is not exactly what Hawking intended to do. But in 1980 Hawking's fears came true when Bekenstein stated: “Having started as a vocal critic of the generalized second law, Hawking became the person who made it fully consistent with the gedanken experiments” \cite{Bekenstein 1980}, p. 29. 

A suggestion has been made that Hawking radiation should more appropriately be called Bekenstein-Hawking radiation, but Bekenstein himself rejected this. He said: “The entropy of a black hole is called Bekenstein-Hawking entropy, which I think is fine. I wrote it down first, Hawking found the numerical value of the constant, so together we found the formula as it is today. The radiation was really Hawking’s work. I had no idea how a black hole could radiate. Hawking brought that out very clearly. So that should be called Hawking radiation” \cite{Clark}. 

\section*{Acknowledgement}

I still remember the inspiring conversations I had in 2005 with Sir Roger Penrose in Israel on Einstein, general relativity and black holes. I would like to thank Juan Maldacena for the historical comments he gave me during our exchange of emails on black holes. I am grateful to Gil Kalai for discussions and helpful comments. Finally, I would like to thank Yaron Sheffer for pointing out an error in an equation in the first version of this paper.


\begin{thebibliography}{53}

\bibitem[1]{Penrose5} An, D., Meissner, K. A., Nurowski, P. and Penrose, R. (2018). "Apparent evidence for Hawking points in the CMB Sky." \emph{Monthly Notices of the Royal Astronomical Society} 495, pp. 3403–3408. 

\bibitem[2]{Bain} Bain, Jonathan (2018). "Black Hole Thermodynamics Part II." Lectures NYU.

\bibitem[3]{BCH} Bardeen, J. M., Carter, B., Hawking, S. W. (1973). “The Four Laws of Black Hole Mechanics.” \emph{Communications in Mathematical Physics} 31, pp. 161-170. 

\bibitem[4]{Bekenstein 1972} Bekenstein, J. D. (1972). “Black Holes and the Second Law.” \emph{Lettere al Nuovo Cimento} 4, pp. 737-740.

\bibitem[5]{Bekenstein 1973} Bekenstein, J. D. (1973). “Black Holes and Entropy.” \emph{Physical Review D}7, pp. 2333-2346.

\bibitem[6]{Bekenstein 1974} Bekenstein, J. D. (1974). “Generalized second law of thermodynamics in black-hole physics.” \emph{Physical Review D}9, pp. 3292-3300.

\bibitem[7]{Bekenstein 1975} Bekenstein, J. D. (1975). “Statistical black-hole thermodynamics.” \emph{Physical Review D}12, pp. 3077-3085.

\bibitem[8]{Bekenstein 1980} Bekenstein, J. D. (1980).“Black Hole Thermodynamics.” \emph{Physics Today}, January, pp. 24-31.

\bibitem[9]{Bekenstein 1981} Bekenstein, J. D. (1981). “Universal upper bound on the entropy-to-energy ratio for bounded systems”. \emph{Physical Review D} 23, pp. 287-298.

\bibitem[10]{Bekenstein 1983} Bekenstein, J. D. (1983). “Entropy bounds and the second law for black holes”. \emph{Physical Review D} 27, pp. 2262-2270.

\bibitem[11]{Bekenstein 1993} Bekenstein, J. D. (1993). “How Fast Does Information Leak Out from a Black Hole?” \emph{Physical Review Letters} 70, pp. 3680-3683.

\bibitem[12]{Bekenstein 1994} Bekenstein, J. D. (1994). “Entropy bounds and black hole remnants.” \emph{Contemporary Physics} 49, pp. 1912-1921.

\bibitem[13]{Bekenstein 2001} Bekenstein, J. D. (2001). “The Limits of Information.” \emph{Studies in History and Philosophy of Modern Physics} 32, pp. 511–524.

\bibitem[14]{Bekenstein 2002} Bekenstein, J. D. (2002). “Quantum Information and Quantum Black Holes.” In Bergmann P.G. and de Sabbata, V.(eds.) \emph{Advances in the Interplay between Quantum and Gravity Physics}, Netherlands: Kluwer academic Publishers, pp. 1-26.

\bibitem[15]{Bekenstein 2004} Bekenstein, J. D. (2004). “Black holes and information theory.” \emph{ Contemporary Physics} 45, pp. 31-43.

\bibitem[16] {Brillouin} Brillouin, L. (1959). \emph{La science et la théorie de l'information}. Sceaux: Jacques Gabay Masson. 

\bibitem[17]{Bousso} Bousso, R (2018). "Black Hole Entropy and the Bekenstein Bound." In Brink, L., Mukhanov, V.F., Rabinovici, E., Phua,K. K. (eds.). \emph{Jacob Bekenstein: The Conservative Revolutionary}. MA, US: World Scientific, pp. 139-158.

\bibitem[18]{Carter} Carter, B. (1971). “Axisymmetric Black Hole Has Only Two Degrees of Freedom.” \emph{Physical Review Letters} 26, pp. 331-333.

\bibitem[19]{Christodoulou} Christodoulou, D. (1970). “Reversib1e and Irreversible Transformations in Black-Hole Physics.” \emph{Physical Review Letters} 25, pp. 1596-1597.

\bibitem[20]{Clark} Clark, Stuart (2018). “A Brief History of Stephen Hawking: A Legacy of Paradox.” \emph{New Scientist}, March 14.

\bibitem[21]{Feynman} Feynman, R (1949). "The Theory of Positrons." \emph{Physical Review}, 76, pp. 749-759.

\bibitem[22]{Hawking and Penrose} Hawking, S. W. and Penrose, R. (1970). “The Singularities of Gravitational Collapse and Cosmology.” \emph{Proceedings of the Royal Society of London A} 314, pp. 529-548.

\bibitem[23]{Hawking 1971} Hawking, S. W. (1971). “Gravitational Radiation from Colliding Black Holes.” \emph{Physical Review Letters} 26, pp. 1344-1346.

\bibitem[24]{Hawking 1973} Hawking S. W. (1973). “The Analogy between black hole mechanics and thermodynamics.” \emph{Annals of the New York Academy of Sciences} 224, pp. 268-271.

\bibitem[25]{Hawking 1974} Hawking, S. W. (1974). “Black hole explosions?” \emph{Nature} 248, pp. 30-31.

\bibitem[26]{Hawking 1975} Hawking, S. W. (1976). “Black holes and thermodynamics." \emph{Physical Review D} 13, pp. 191-197.

\bibitem[27]{Hawking 1976} Hawking, S. W. (1976). “Breakdown of predictability in gravitational collapse.” \emph{Physical Review D} 14, pp. 2460-2473.

\bibitem[28]{Hawking 1977}Hawking, S. W. (1977). “The Quantum Mechanics of Black Holes.” \emph{Scientific American} 236, pp. 34-42.

\bibitem[29]{Hawking 1978}Hawking, S. W. (1978). “Black Holes and Unpredictability.” \emph{Physics Bulletin,} January, pp. 23-24.

\bibitem[30]{Hawking 1988} Hawking, S. W. (1988). \emph{A brief History of Time. From the Big Bang to Black Holes.} Introduction by Carl Sagan. New-York: Bantam Books.

\bibitem[31]{Hawking 2003} Hawking, S. W. (2003). "Sixty Years in a Nutshell." In Gibbons, G.W., Shellard, E. P. S and Rankin, S.J. (eds.) \emph{The Future of Theoretical Physics and Cosmology: Celebrating Stephen Hawking's 60th Birthday}. Cambridge: Cambridge University Press. pp. 105-118.

\bibitem[32]{Hawking 2015} Hawking, S. W. (2015). “The Information Paradox for Black Holes.” \emph{ArXiv} 1509:01147v1. [hep-th].

\bibitem[33]{Israel} Israel, W. (1967). “Event Horizons in Static Vacuum Space-Times.” \emph{Physical Review} 164, pp. 1776-1779.

\bibitem[34]{Leonhardt} Leonhardt U. and Philbin T. G. (2008). “The Case for Artificial Black Holes.” \emph{Philosophical Transactions: Mathematical, Physical and Engineering Sciences} 366, pp. 2851-2857.

\bibitem[35]{Nordmann} Nordmann, C. (1922). "Einstein expose et discute sa théorie." \emph{Revue des Deux Mondes} 9, pp. 130-166.

\bibitem[36]{Overbye} Overbye, D. (1979). "The Wizard of Space and Time." \emph{Omni}, February, pp. 45-46, pp. 104-107. 

\bibitem[37]{Penrose3} Penrose, R. (1964). "Conformal treatment of infinity." In deWitt, B. and deWitt, C. (eds.), \emph{Relativity, groups and topology}. Gordon and Breach, New York, pp. 565–584. 

\bibitem[38]{Penrose} Penrose, R. (1969). “Gravitational Collapse: The Role of General Relativity.” \emph{Rivista del Nuovo Cimento, Numero Speziale} 1, pp. 252-275; \emph{General Relativity and Gravitation} 34, 2002, pp. 1141-1165.

\bibitem[39]{Penrose4} Penrose, R. (2018). "The Big Bang and its Dark-Matter Content: Whence, Whither, and Wherefore." \emph{Foundations of Physics} 48, pp. 1177–1190. 

\bibitem[40]{Penrose6} Penrose, R. (2019). "Are we Seeing Hawking Points in the Microwave Sky?" Lecture at the Institut Astrophysique de Paris, CNRS. 

\bibitem[41]{Penrose2} Penrose, R. and Floyd, R. M (1971). “Extraction of Rotational Energy from a Black Hole.” \emph{Nature Physical Science} 229, pp. 177-179.

\bibitem[42]{Rovelli} Rovelli, C. E. (1996). “Relational Quantum Mechanics.” \emph{International Journal of Theoretical Physics} 35, pp. 1637-1678.

\bibitem[43]{Shannon} Shannon, C. E. (1948). “A Mathematical Theory of Communication.” \emph{The Bell System Technical Journal} 27, pp. 379-423.

\bibitem[44]{Szilard} Szilard, L. (1929). "Über die Entropieverminderung in einem thermodynamischen System bei Eingriffen intelligenter Wesen." \emph{Zeitschrift für Physik } 53, pp. 840–856; “On the decrease of entropy in a thermodynamic system by the intervention of intelligent beings.” \emph{Behavioral Science} 9, pp. 301-310, 1964. 

\bibitem[45]{Thorne} Thorne, K. S. (1994). \emph{Black Holes and Time Warps. Einstein's Outrageous Legacy}. Forward by S. Hawking. New York: W. W. Norton.

\bibitem[46]{Unruh} Unruh, W. G. (1981). “Experimental Black Hole Evaporation?” \emph{Physical Review Letters} 46, pp. 1351-1353.

\bibitem[47]{Unruh2} Unruh, W. G. (2001). "Black hole, dumb holes, and entropy." In Callender C. and Huggett N., \emph{Physics meets philosophy at the Planck scale. Contemporary theories in quantum gravity}. New York: Cambridge University Press, pp. 152-175. 

\bibitem[48]{Unruh3} Unruh, W. G. (2018). "Black Holes in the Laboratory." \emph{Bekenstein Memorial Lecture}, Hebrew University of Jerusalem. 

\bibitem[49]{Unruh and Wald} Unruh, W. G. and Wald R. M. (1981). "Acceleration radiation and the generalized second law of thermodynamics." \emph{Physical Review D} 25, pp. 942-958. 

\bibitem[50]{Visser} Visser, M. (2003). “Essential and inessential features of Hawking radiation.” \emph{International Journal of Modern Physics D} 12, p. 649–661.

\bibitem[51]{Wald} Wald, R. M. (2020). “Jacob Bekenstein and the Development of Black Hole Thermodynamics.” In Brink, L., Mukhanov, V.F., Rabinovici, E., Phua, K. K. (eds.). \emph{Jacob Bekenstein: The Conservative Revolutionary}. MA, US: World Scientific, pp. 3-10. 

\bibitem[52]{Wheeler 1981} Wheeler, J. A. (1981). “The Lesson of the Black Hole.” \emph{Proceedings of the American Philosophical Society} 125, pp. 25-37. 

\bibitem[53]{Wheeler 2008} Wheeler, J. A. (2008). “Entropy of a black hole. Bekenstein, Stephen Hawking.” \emph{Web of Stories}. 

\end{thebibliography}
\end{document}